\documentclass{article}%
\usepackage{amsmath}
\usepackage{amsfonts}
\usepackage{amssymb}
\usepackage{graphicx}%
\providecommand{\U}[1]{\protect\rule{.1in}{.1in}}

\begin{document}

\title{The Basics of Information Geometry\thanks{Invited tutorial presented at MaxEnt
2014, the 34th International Workshop on Bayesian Inference and Maximum
Entropy Methods in Science and Engineering (September 21--26, 2014, Amboise,
France). }}
\author{Ariel Caticha\\{\small Physics Department, University at Albany-SUNY, Albany, NY 12222, USA.}}
\date{}
\maketitle

\begin{abstract}
To what extent can we distinguish one probability distribution from another?
Are there quantitative measures of distinguishability?\ The goal of this
tutorial is to approach such questions by introducing the notion of the
\textquotedblleft distance\textquotedblright\ between two probability
distributions and exploring some basic ideas of such an \textquotedblleft
information geometry\textquotedblright.

\end{abstract}

\begin{description}
\item[Einstein, 1949:] \textquotedblleft\lbrack The basic ideas of General
Relativity were in place] ... \emph{in 1908. Why were another seven years
required for the construction of the general theory of relativity? The main
reason lies in the fact that it is not so easy to free oneself from the idea
that coordinates must have an immediate metrical meaning.\textquotedblright%
\ }\cite{Einstein 1949}
\end{description}

\section{Introduction}

A main concern of any theory of inference is the problem of updating
probabilities when new information becomes available. We want to pick a
probability distribution from a set of candidates and this immediately raises
many questions. What if we had picked a neighboring distribution? What
difference would it make?\ What makes two distributions similar? To what
extent can we distinguish one distribution from another? Are there
quantitative measures of distinguishability?\ The goal of this tutorial is to
address such questions by introducing methods of geometry. More specifically
the goal will be to introduce a notion of \textquotedblleft
distance\textquotedblright\ between two probability distributions.

A parametric family of probability distributions is a set of distributions
$p_{\theta}(x)$ labeled by parameters $\theta=(\theta^{1}\ldots\theta^{n})$.
Such a family forms a \emph{statistical manifold}, namely, a space in which
each point, labeled by coordinates $\theta$, represents a probability
distribution $p_{\theta}(x)$. Generic manifolds do not come with an intrinsic
notion of distance; such additional structure has to be supplied separately in
the form of a metric tensor. Statistical manifolds are, however, an exception.
One of the main goals of this chapter is to show that statistical manifolds
possess a uniquely natural notion of distance --- the so-called information
metric. This metric is not an optional feature; it is inevitable.
\emph{Geometry is intrinsic to the structure of statistical manifolds.}

The distance $d\ell$ between two neighboring points $\theta$ and
$\theta+d\theta$ is given by Pythagoras' theorem which, written in terms of a
metric tensor $g_{ab}$, is\footnote{The use of superscripts rather than
subscripts for the indices labelling coordinates is a standard and very
convenient notational convention in differential geometry. We adopt the
standard convention of summing over repeated indices, for example,
$g_{ab}f^{ab}=%
{\textstyle\sum\nolimits_{a}}
{\textstyle\sum\nolimits_{b}}
g_{ab}f^{ab}$.}
\begin{equation}
d\ell^{2}=g_{ab}d\theta^{a}d\theta^{b}~.
\end{equation}
The singular importance of the metric tensor $g_{ab}$ derives from a theorem
due to N. \v{C}encov that states that the metric $g_{ab}$ on the manifold of
probability distributions is essentially unique: up to an overall scale factor
there is only one metric that takes into account the fact that these are not
distances between simple structureless dots but distances between probability
distributions. \cite{Cencov 1981}

We will not develop the subject in all its possibilities\footnote{For a more
extensive treatment see \cite{Amari 1985}\cite{Amari Nagaoka 2000}. Here we
follow closely the presentation in \cite{Caticha 2012}.} but we do wish to
emphasize one specific result. Having a notion of distance means we have a
notion of volume and this in turn implies that \emph{there is a unique and
objective notion of a distribution that is uniform over the space of
parameters} --- equal volumes are assigned equal probabilities. Whether such
uniform distributions are maximally non-informative, or whether they define
ignorance, or whether they reflect the actual prior beliefs of any rational
agent, are all important issues but they are quite beside the specific point
that we want to make, namely, that they are uniform --- and this is not a
matter of subjective judgment but of objective mathematical proof.

\section{Examples of statistical manifolds}

An $n$-dimensional manifold $\mathcal{M}$ is a smooth, possibly curved, space
that is locally like $\mathcal{R}^{n}$. What this means is that one can set up
a coordinate frame (that is a map $\mathcal{M}\rightarrow\mathcal{R}^{n}$) so
that each point $\theta\in\mathcal{M}$ is identified or labelled by its
coordinates, $\theta=(\theta^{1}\ldots\theta^{n})$. A statistical manifold is
a manifold in which each point $\theta$ represents a probability distribution
$p_{\theta}(x)$. As we shall later see, a very convenient notation is
$p_{\theta}(x)=p(x|\theta)$. Here are some examples:

The \textbf{multinomial distributions}\emph{ }are given by
\begin{equation}
p(\{n_{i}\}|\theta)=\frac{N!}{n_{1}!n_{2}!\ldots n_{m}!}(\theta^{1})^{n_{1}%
}(\theta^{2})^{n_{2}}\ldots(\theta^{m})^{n_{m}}~, \label{multinomial}%
\end{equation}
where $\theta=(\theta^{1},\theta^{2}\ldots\theta^{m})$, $N=%
{\textstyle\sum\nolimits_{i=1}^{m}}
n_{i}$ and $%
{\textstyle\sum\nolimits_{i=1}^{m}}
\theta^{i}=1$. They form a statistical manifold of dimension $(m-1)$ called a
simplex, $S_{m-1}$. The parameters $\theta=(\theta^{1},\theta^{2}\ldots
\theta^{m})$ are a convenient choice of coordinates.

The \textbf{multivariate Gaussian distributions}\emph{ }with means $\mu^{a}$,
$a=1\ldots n$, and variance $\sigma^{2}$,
\begin{equation}
p(x|\mu,\sigma)=\frac{1}{(2\pi\sigma^{2})^{n/2}}\exp-\frac{1}{2\sigma^{2}}%
{\textstyle\sum\limits_{a=1}^{n}}
(x^{a}-\mu^{a})^{2}~,
\end{equation}
form an $(n+1)$-dimensional statistical manifold with coordinates $\theta
=(\mu^{1},\ldots,\mu^{n},\sigma^{2})$.

The \textbf{canonical distributions},
\begin{equation}
p(i|F)=\frac{1}{Z}e^{-\lambda_{k}f_{i}^{k}}~,
\end{equation}
are derived by maximizing the Shannon entropy $S[p]$ subject to constraints on
the expected values of $n$ functions $f_{i}^{k}=f^{k}(x_{i})$ labeled by
superscripts $k=1,2,\ldots n$,%
\begin{equation}
\left\langle f^{k}\right\rangle =%
{\textstyle\sum\limits_{i}}
p_{i}f_{i}^{k}=F^{k}~.
\end{equation}
They form an $n$-dimensional statistical manifold. As coordinates we can
either use the expected values $F=(F^{1}\ldots F^{n})$ or, equivalently, the
Lagrange multipliers, $\lambda=(\lambda_{1}\ldots\lambda_{n})$.

\section{Distance and volume in curved spaces}

The basic intuition behind differential geometry derives from the observation
that curved spaces are locally flat: curvature effects can be neglected
provided one remains within a sufficiently small region. The idea then is
rather simple: within the close vicinity of any point $x$ we can always
transform from the original coordinates $x^{a}$ to new coordinates $\hat
{x}^{\alpha}=\hat{x}^{\alpha}(x^{1}\ldots x^{n})$ that we \emph{declare} to be
locally Cartesian (here denoted with a hat and with Greek superscripts,
$\hat{x}^{\alpha}$). An infinitesimal displacement is given by
\begin{equation}
d\hat{x}^{\alpha}=X_{a}^{\alpha}\,dx^{a}\quad\text{where}\quad X_{a}^{\alpha
}=\frac{\partial\hat{x}^{\alpha}}{\partial x^{a}} \label{Cartesian dx}%
\end{equation}
and the corresponding infinitesimal distance can be computed using Pythagoras
theorem,
\begin{equation}
d\ell^{2}=\delta_{\alpha\beta}d\hat{x}^{\alpha}d\hat{x}^{\beta}~.
\end{equation}
Changing back to the original frame
\begin{equation}
d\ell^{2}=\delta_{\alpha\beta}d\hat{x}^{\alpha}d\hat{x}^{\beta}=\delta
_{\alpha\beta}X_{a}^{\alpha}\,X_{b}^{\beta}\,dx^{a}dx^{b}~.
\end{equation}
Defining the quantities
\begin{equation}
g_{ab}\equiv\delta_{\alpha\beta}X_{a}^{\alpha}\,X_{b}^{\beta}~,
\label{metric transf a}%
\end{equation}
we can write the infinitesimal Pythagoras theorem in generic coordinates
$x^{a}$ as%
\begin{equation}
d\ell^{2}=g_{ab}dx^{a}dx^{b}~. \label{metric dist}%
\end{equation}
The quantities $g_{ab}$ are the components of the metric tensor. One can
easily check that under a coordinate transformation $g_{ab}$ transforms
according to
\begin{equation}
g_{ab}=X_{a}^{a^{\prime}}X_{a}^{b^{\prime}}g_{a^{\prime}b^{\prime}}%
\quad\text{where}\quad X_{a}^{a^{\prime}}=\frac{\partial x^{a^{\prime}}%
}{\partial x^{a}}\,, \label{metric transf b}%
\end{equation}
so that the infinitesimal distance $d\ell$ is independent of the choice of coordinates.

To find the finite length between two points along a curve $x(\lambda)$ one
integrates along the curve,
\begin{equation}
\ell=\int_{\lambda_{1}}^{\lambda_{2}}d\ell=\int_{\lambda_{1}}^{\lambda_{2}%
}\left(  g_{ab}\frac{dx^{a}}{d\lambda}\frac{dx^{b}}{d\lambda}\right)
^{1/2}d\lambda~. \label{length}%
\end{equation}
Once we have a measure of distance we can also measure angles, areas, volumes
and all sorts of other geometrical quantities. To find an expression for the
$n$-dimensional volume element $dV_{n}$ we use the same trick as before:
transform to locally Cartesian coordinates so that the volume element is
simply given by the product
\begin{equation}
dV_{n}=d\hat{x}^{1}d\hat{x}^{2}\ldots d\hat{x}^{n}~,
\end{equation}
and then transform back to the original coordinates $x^{a}$ using
eq.(\ref{Cartesian dx}),
\begin{equation}
dV_{n}=\left\vert \frac{\partial\hat{x}}{\partial x}\right\vert dx^{1}%
dx^{2}\ldots dx^{n}=\left\vert \det X_{a}^{\alpha}\right\vert d^{n}x~.
\end{equation}
This is the volume we seek written in terms of the coordinates $x^{a}$ but we
still have to calculate the Jacobian of the transformation, $\left\vert
\partial\hat{x}/\partial x\right\vert =\left\vert \det X_{a}^{\alpha
}\right\vert $. The transformation of the metric from its Euclidean form
$\delta_{\alpha\beta}$ to $g_{ab}$, eq.(\ref{metric transf a}), is the product
of three matrices. Taking the determinant we get
\begin{equation}
g\equiv\det(g_{ab})=\left[  \det X_{a}^{\alpha}\right]  ^{2}~,
\end{equation}
so that
\begin{equation}
\left\vert \det\left(  X_{a}^{\alpha}\right)  \right\vert =g^{1/2}~.
\end{equation}
We have succeeded in expressing the volume element in terms of the metric
$g_{ab}(x)$ in the original coordinates $x^{a}$. The answer is
\begin{equation}
dV_{n}=g^{1/2}(x)d^{n}x~.
\end{equation}
The volume of any extended region on the manifold is
\begin{equation}
V_{n}=\int dV_{n}=\int g^{1/2}(x)d^{n}x~.~
\end{equation}

\noindent\textbf{Example:} A uniform distribution over such a curved manifold
is one which assigns equal probabilities to equal volumes,
\begin{equation}
p(x)d^{n}x\propto g^{1/2}(x)d^{n}x~.
\end{equation}

\noindent\textbf{Example:} For Euclidean space in spherical coordinates
$(r,\theta,\phi)$,
\begin{equation}
d\ell^{2}=dr^{2}+r^{2}d\theta^{2}+r^{2}\sin^{2}\theta d\phi^{2}~,
\end{equation}
and the volume element is the familiar expression
\begin{equation}
dV=g^{1/2}drd\theta d\phi=r^{2}\sin\theta\,drd\theta d\phi~.
\end{equation}

\section{Two derivations of the information metric}

The distance $d\ell$ between two neighboring distributions $p(x|\theta)$ and
$p(x|\theta+d\theta)$ or, equivalently, between the two points $\theta$ and
$\theta+d\theta$, is given by the metric $g_{ab}$. Our goal is to compute the
tensor $g_{ab}$ corresponding to $p(x|\theta)$. We give a couple of
derivations which illuminate the meaning of the information metric, its
interpretation, and ultimately, how it is to be used. Other derivations based
on asymptotic inference are given in \cite{Wootters 1981} and
\cite{Balasubramanian 1997}.

At this point a word of caution (and encouragement) might be called for. Of
course it is possible to be confronted with sufficiently singular families of
distributions that are not smooth manifolds and studying their geometry might
seem a hopeless enterprise. Should we give up on geometry? No. The fact that
statistical manifolds can have complicated geometries does not detract from
the value of the methods of information geometry any more than the existence
of surfaces with rugged geometries detracts from the general value of geometry itself.

\subsubsection*{Derivation from distinguishability}

We seek a quantitative measure of the extent that two distributions
$p(x|\theta)$ and $p(x|\theta+d\theta)$ can be distinguished. The following
argument is intuitively appealing. \cite{Rao 1945}\cite{Atkinson Mitchell
1981} The advantage of this approach is that it clarifies the interpretation
--- \emph{the metric measures distinguishability}. Consider the relative
difference,
\begin{equation}
\Delta=\frac{p(x|\theta+d\theta)-p(x|\theta)}{p(x|\theta)}=\frac{\partial\log
p(x|\theta)}{\partial\theta^{{}a}}\,d\theta^{{}a}.
\end{equation}
The expected value of the relative difference, $\langle\Delta\rangle$, might
seem a good candidate, but it does not work because it vanishes identically,
\begin{equation}
\langle\Delta\rangle=\int dx\,p(x|\theta)\,\frac{\partial\log p(x|\theta
)}{\partial\theta^{{}a}}\,d\theta^{{}a}=d\theta^{{}a}\,\frac{\partial
}{\partial\theta^{{}a}}\int dx\,p(x|\theta)=0.
\end{equation}
(Depending on the problem the symbol $%
{\textstyle\int}
dx$ may represent either discrete sums or integrals over one or more
dimensions; its meaning should be clear from the context.) However, the
variance does not vanish,
\begin{equation}
d\ell^{2}=\langle\Delta^{2}\rangle=\int dx\,p(x|\theta)\,\frac{\partial\log
p(x|\theta)}{\partial\theta^{{}a}}\,\frac{\partial\log p(x|\theta)}%
{\partial\theta^{{}b}}\,d\theta^{{}a}d\theta^{{}b}\,\,.
\end{equation}
This is the measure of distinguishability we seek; a small value of $d\ell
^{2}$ means that the relative difference $\Delta$ is small and the points
$\theta$ and $\theta+d\theta$ are difficult to distinguish. It suggests
introducing the matrix $g_{ab}$
\begin{equation}
g_{ab}(\theta)\overset{\text{def}}{=}\int dx\,p(x|\theta)\,\frac{\partial\log
p(x|\theta)}{\partial\theta^{{}a}}\,\frac{\partial\log p(x|\theta)}%
{\partial\theta^{{}b}} \label{info metric a}%
\end{equation}
called the Fisher information \emph{matrix} \cite{Fisher 1925}, so that
\begin{equation}
d\ell^{2}=g_{ab}\,d\theta^{{}a}d\theta^{{}b}\,\,. \label{info metric b}%
\end{equation}

Up to now no notion of distance has been introduced. Normally one says that
the reason it is difficult to distinguish two points in say, the three
dimensional space we seem to inhabit, is that they happen to be too close
together. It is tempting to invert this intuition and assert that two points
$\theta$ and $\theta+d\theta$ are \emph{close} together whenever they are
difficult to distinguish. Furthermore, being a variance, the quantity
$d\ell^{2}=\langle\Delta^{2}\rangle$ is positive and vanishes only when
$d\theta$ vanishes. Thus, it is natural to introduce distance by interpreting
$g_{ab}$ as the metric tensor of a Riemannian space. \cite{Rao 1945} This is
the \emph{information metric}. The recognition by Rao that $g_{ab}$ is a
metric in the space of probability distributions gave rise to the subject of
information geometry \cite{Amari 1985}, namely, the application of geometrical
methods to problems in inference and in information theory.

The coordinates $\theta$ are quite arbitrary; one can freely relabel the
points in the manifold. It is then easy to check that $g_{ab}$ are the
components of a tensor and that the distance $d\ell^{2}$ is an invariant, a
scalar under coordinate transformations. Indeed, the transformation
\begin{equation}
\theta^{a^{\prime}}=f^{a^{\prime}}(\theta^{1}\ldots\theta^{n})
\end{equation}
leads to
\begin{equation}
d\theta^{a}=\frac{\partial\theta^{a}}{\partial\theta^{{}a^{\prime}}}%
d\theta^{a^{\prime}}\quad\text{and}\quad\frac{\partial}{\partial\theta^{{}a}%
}=\frac{\partial\theta^{a^{\prime}}}{\partial\theta^{{}a}}\frac{\partial
}{\partial\theta^{{}a^{\prime}}}%
\end{equation}
so that, substituting into eq.(\ref{info metric a}),
\begin{equation}
g_{ab}=\frac{\partial\theta^{a^{\prime}}}{\partial\theta^{{}a}}\frac
{\partial\theta^{b^{\prime}}}{\partial\theta^{{}b}}g_{a^{\prime}b^{\prime}}%
\end{equation}

\subsubsection*{Derivation from relative entropy}

Elsewhere we argued for the concept of relative entropy $S[p,q]$ as a tool for
updating probabilities from a prior $q$ to a posterior $p$ when new
information in the form of constraints becomes available. (For a detailed
development of the Method of Maximum Entropy see \cite{Caticha 2012} and
references therein.) The idea is to use $S[p,q]$ to rank those distributions
$p$ relative to $q$ so that the preferred posterior is that which maximizes
$S[p,q]$ subject to the constraints. The functional form of $S[p,q]$ is
derived from very conservative design criteria that recognize the value of
information: what has been learned in the past is valuable and should not be
disregarded unless rendered obsolete by new information. This is expressed as
a \emph{Principle of Minimal Updating}: beliefs should be revised only to the
extent required by the new evidence. According to this interpretation those
distributions $p$ that have higher entropy $S[p,q]$ are \emph{closer} to $q$
in the sense that they reflect a less drastic revision of our beliefs.

The term `closer' is very suggestive but it can also be dangerously
misleading. On one hand, it suggests there is a connection between entropy and
geometry. As shown below, such a connection does, indeed, exist. On the other
hand, it might tempt us to \emph{identify} $S[p,q]$ with distance which is,
obviously, incorrect: $S[p,q]$ is not symmetric, $S[p,q]\neq S[q,p]$, and
therefore it cannot be a distance. There is a relation between entropy and
distance but the relation is not one of identity.

In curved spaces the distance between two points $p$\ and $q$ is the length of
the shortest curve that joins them and the length $\ell$ of a curve,
eq.(\ref{length}), is the sum of \emph{local} infinitesimal lengths $d\ell$
lying between $p$ and $q$. On the other hand, the entropy $S[p,q]$ is a
\emph{non-local} concept. It makes no reference to any points other than $p$
and $q$. Thus, the relation between entropy and distance, if there is any all,
must be a relation between two infinitesimally close distributions $q$ and
$p=q+dq$. Only in this way can we define a distance without referring to
points between $p$ and $q$. (See also \cite{Rodriguez 1989}.)

Consider the entropy of one distribution $p(x|\theta^{\prime})$ relative to
another $p(x|\theta)$,
\begin{equation}
S(\theta^{\prime},\theta)=-\int dx\,p(x|\theta^{\prime})\log\frac
{p(x|\theta^{\prime})}{p(x|\theta)}~.
\end{equation}
We study how this entropy varies when $\theta^{\prime}=\theta+d\theta$ is in
the close vicinity of a given $\theta$. It is easy to check -- recall the
Gibbs inequality, $S(\theta^{\prime},\theta)\leq0$, with equality if and only
if $\theta^{\prime}=\theta$ --- that the entropy $S(\theta^{\prime},\theta)$
attains an absolute maximum at $\theta^{\prime}=\theta$ . Therefore, the first
nonvanishing term in the Taylor expansion about $\theta$ is second order in
$d\theta$%
\begin{equation}
S(\theta+d\theta,\theta)=\left.  \frac{1}{2}\frac{\partial^{2}S(\theta
^{\prime},\theta)}{\partial\theta^{\prime a}\partial\theta^{\prime b}%
}\right\vert _{\theta^{\prime}=\theta}d\theta^{a}d\theta^{b}+\ldots\leq0~,
\end{equation}
which suggests defining a distance $d\ell$ by
\begin{equation}
S(\theta+d\theta,\theta)=-\frac{1}{2}d\ell^{2}~.
\end{equation}
A straightforward calculation of the second derivative gives the information
metric,
\begin{equation}
\left.  -\frac{\partial S(\theta^{\prime},\theta)}{\partial\theta^{\prime
a}\partial\theta^{\prime b}}\right\vert _{\theta^{\prime}=\theta}=\int
dx\,p(x|\theta)\frac{\partial\log p(x|\theta)}{\partial\theta^{a}}%
\frac{\partial\log p(x|\theta)}{\partial\theta^{b}}=g_{ab}~.
\label{info metric i}%
\end{equation}

\section{Uniqueness of the information metric}

A most remarkable fact about the information metric is that it is essentially
unique: except for a constant scale factor it is the only Riemannian metric
that adequately takes into account the nature of the points of a statistical
manifold, namely, that these points represent probability distributions, that
they are not \textquotedblleft structureless\textquotedblright. This theorem
was first proved by N. \v{C}encov within the framework of category theory
\cite{Cencov 1981}; later Campbell gave an alternative proof that relies on
the notion of Markov mappings. \cite{Campbell 1986} Here I will describe
Campbell's basic idea in the context of a simple example.

We can use binomial distributions to analyze the tossing of a coin (with
probabilites $p($heads$)=\theta$ and $p($tails$)=1-\theta$). We can also use
binomials to describe the throwing of a special die. For example, suppose that
the die is loaded with equal probabilities for three faces, $p_{1}=p_{2}%
=p_{3}=\theta/3$, and equal probabilities for the other three faces,
$p_{4}=p_{5}=p_{6}=(1-\theta)/3$. Then we use a binomial distribution to
describe the coarse outcomes low$\,=\{1,2,3\}$ or high$\,=\{4,5,6\}$ with
probabilities $\theta$ and $1-\theta$. This amounts to mapping the space of
coin distributions to a subspace of the space of die distributions.

The embedding of the statistical manifold of $n=2$ binomials, which is a
simplex $\mathcal{S}_{1}$ of dimension one, into a subspace of the statistical
manifold of $n=6$ multinomials, which is a simplex $\mathcal{S}_{5}$ of
dimension five, is called a Markov mapping.

Having introduced the notion of Markov mappings we can now state the basic
idea behind Campbell's argument: whether we talk about heads/tails outcomes in
coins or we talk about low/high outcomes in dice, binomials are binomials.
Whatever geometrical relations are assigned to distributions in $\mathcal{S}%
_{1}$, exactly the same geometrical relations should be assigned to the
distributions in the corresponding subspace of $\mathcal{S}_{5}$. Therefore,
these Markov mappings are not just embeddings, they are congruent\ embeddings
--- distances between distributions in $\mathcal{S}_{1}$ should match the
distances between the corresponding images in $\mathcal{S}_{5}$.

Now for the punch line: the goal is to find the Riemannian metrics that are
invariant under Markov mappings. It is easy to see why imposing such
invariance is extremely restrictive: The fact that distances computed in
$\mathcal{S}_{1}$ must agree with distances computed in subspaces of
$\mathcal{S}_{5}$ introduces a constraint on the allowed metric tensors; but
we can always embed $\mathcal{S}_{1}$ and $\mathcal{S}_{5}$ in spaces of
larger and larger dimension which leads to more and more constraints. It could
very well have happened that no Riemannian metric survives such restrictive
conditions; it is quite remarkable that some do survive and it is even more
remarkable that (up to an uninteresting scale factor) the surviving Riemannian
metric is unique. Details of the proof are given in \cite{Caticha 2012}.

\section{The metric for some common distributions}

The statistical manifold of \textbf{multinomial distributions},
\begin{equation}
P_{N}\left(  n|\theta\right)  =\frac{N!}{n_{1}!\ldots n_{m}!}\theta_{1}%
^{n_{1}}\ldots\theta_{m}^{n_{m}}~,
\end{equation}
where
\begin{equation}
n=(n_{1}\ldots n_{m})\quad\text{with}\quad%
{\textstyle\sum\limits_{i=1}^{m}}
n_{i}=N\quad\text{and}\quad%
{\textstyle\sum\limits_{i=1}^{m}}
\theta_{i}=1\,,
\end{equation}
is the simplex $\mathcal{S}_{m-1}$. The metric is given by
eq.(\ref{info metric a}),
\begin{equation}
g_{ij}=%
{\textstyle\sum\limits_{n}}
P_{N}\frac{\partial\log P_{N}}{\partial\theta_{i}}\frac{\partial\log P_{N}%
}{\partial\theta_{j}}\quad\text{where}\quad1\leq i,j\leq m-1~.
\end{equation}
The result is
\begin{equation}
g_{ij}=\left\langle (\frac{n_{i}}{\theta_{i}}-\frac{n_{m}}{\theta_{m}}%
)(\frac{n_{j}}{\theta_{j}}-\frac{n_{m}}{\theta_{m}})\right\rangle =\frac
{N}{\theta_{i}}\delta_{ij}+\frac{N}{\theta_{m}}~,
\end{equation}
where $1\leq i$, $j\leq m-1$. A somewhat simpler expression can be obtained
writing $d\theta_{m}=-%
{\textstyle\sum\nolimits_{i=1}^{m-1}}
d\theta_{i}$ and extending the range of the indices to include $i$, $j=m$. The
result is%
\begin{equation}
d\ell^{2}=%
{\textstyle\sum\limits_{i,j=1}^{m}}
g_{ij}d\theta_{i}d\theta_{j}\quad\text{with}\quad g_{ij}=\frac{N}{\theta_{i}%
}\delta_{ij}~.
\end{equation}
A uniform distribution over the simplex $\mathcal{S}_{m-1}$ assigns equal
probabilities to equal volumes,
\begin{equation}
P(\theta)d^{m-1}\theta\propto g^{1/2}d^{m-1}\theta\quad\text{with}\quad
g=\frac{N^{m-1}}{\theta_{1}\theta_{2}\ldots\theta_{m}}
\label{multinomial uniform}%
\end{equation}
In the particular case of binomial distributions $m=2$ with $\theta_{1}%
=\theta$ and $\theta_{2}=1-\theta$ we get%
\begin{equation}
g=g_{11}=\frac{N}{\theta(1-\theta)}%
\end{equation}
so that the uniform distribution over $\theta$ (with $0<\theta<1$) is
\begin{equation}
P(\theta)d\theta\propto\lbrack\frac{N}{\theta(1-\theta)}]^{1/2}d\theta~.
\end{equation}

\textbf{Canonical distributions:} Let $z$ denote the microstates of a system
(\emph{e.g.}, points in phase space) and let $m(z)$ be the underlying measure
(\emph{e.g.}, a uniform density on phase space). The space of macrostates is a
statistical manifold: each macrostate is a canonical distribution obtained by
maximizing entropy $S[p,m]$ subject to $n$ constraints $\langle f^{a}%
\rangle=F^{a}$ for $a=1\ldots n$, plus normalization,%
\begin{equation}
p(z|F)=\frac{1}{Z(\lambda)}m(z)e^{-\lambda_{a}f^{a}(z)}\quad\text{where}\quad
Z(\lambda)=\int dz\,m(z)e^{-\lambda_{a}f^{a}(z)}~. \label{can1}%
\end{equation}
The set of numbers $F=(F^{1}\ldots F^{n})$ determines one point $p(z|F)$ on
the statistical manifold so we can use the $F^{a}$ as coordinates.

First, here are some useful facts about canonical distributions. The Lagrange
multipliers $\lambda_{a}$ are implicitly determined by%
\begin{equation}
\langle f^{a}\rangle=F^{a}=-\frac{\partial\log Z}{\partial\lambda_{a}}\,,
\label{can2}%
\end{equation}
and it is straightforward to show that a further derivative with respect to
$\lambda_{b}$ yields the covariance matrix,
\begin{equation}
C^{ab}\equiv\langle(f^{a}-F^{a})(f^{b}-F^{b})\rangle=-\frac{\partial F^{a}%
}{\partial\lambda_{b}}~.
\end{equation}
Furthermore, from the chain rule
\begin{equation}
\delta_{a}^{c}=\frac{\partial\lambda_{a}}{\partial\lambda_{c}}=\frac
{\partial\lambda_{a}}{\partial F^{b}}\frac{\partial F^{b}}{\partial\lambda
_{c}}~,
\end{equation}
it follows that the matrix
\begin{equation}
C_{ab}=-\frac{\partial\lambda_{a}}{\partial F^{b}}%
\end{equation}
is the inverse of the covariance matrix, $C_{ab}C^{bc}=\delta_{a}^{c}~.$

The information metric is%
\begin{align}
g_{ab}  &  =\int dz\,p(z|F)\,\frac{\partial\log p(z|F)}{\partial F^{{}a}%
}\,\frac{\partial\log p(z|F)}{\partial F^{{}b}}\nonumber\\
&  =\frac{\partial\lambda_{c}}{\partial F^{a}}\frac{\partial\lambda_{d}%
}{\partial F^{b}}\int dz\,p\,\frac{\partial\log p}{\partial\lambda_{c}}%
\,\frac{\partial\log p}{\partial\lambda_{d}}~.
\end{align}
Using eqs.(\ref{can1}) and (\ref{can2}),
\begin{equation}
\frac{\partial\log p(z|F)}{\partial\lambda_{c}}=F^{c}-f^{c}(z)
\end{equation}
therefore,
\begin{equation}
g_{ab}=C_{ca}C_{db}C^{cd}\implies g_{ab}=C_{ab}~, \label{can3}%
\end{equation}
so that the metric tensor $g_{ab}$ is the inverse of the covariance matrix
$C^{ab}$.

Instead of the expected values $F^{a}$ we could have used the Lagrange
multipliers $\lambda_{a}$ as coordinates. Then the information metric is the
covariance matrix,
\begin{equation}
g^{ab}=\int dz\,p(z|\lambda)\,\frac{\partial\log p(z|\lambda)}{\partial
\lambda_{a}}\,\frac{\partial\log p(z|\lambda)}{\partial\lambda_{b}}=C^{ab}~.
\label{can4}%
\end{equation}
Therefore the distance $d\ell$ between neighboring distributions can written
in either of two equivalent forms,
\begin{equation}
d\ell^{2}=g_{ab}dF^{a}dF^{b}=g^{ab}d\lambda_{a}d\lambda_{b}~. \label{can5}%
\end{equation}
The uniform distribution over the space of macrostates assigns equal
probabilities to equal volumes,
\begin{equation}
P(F)d^{n}F\propto C^{-1/2}d^{n}F\quad\text{or}\quad P^{\prime}(\lambda
)d^{n}\lambda\propto C^{1/2}d^{n}\lambda~,
\end{equation}
where $C=\det C^{ab}$.

\textbf{Gaussian distributions} are a special case of canonical distributions
--- they maximize entropy subject to constraints on mean values and
correlations. Consider Gaussian distributions in $D$ dimensions,
\begin{equation}
p(x|\mu,C)=\frac{c^{1/2}}{(2\pi)^{D/2}}\exp\left[  -\frac{1}{2}C_{ij}%
(x^{i}-\mu^{i})(x^{j}-\mu^{j})\right]  ~,
\end{equation}
where $1\leq i\leq D$, $C_{ij}$ is the inverse of the correlation matrix, and
$c=\det C_{ij}$. The mean values $\mu^{i}$ are $D$ parameters $\mu^{i}$, while
the symmetric $C_{ij}$ matrix is an additional $\frac{1}{2}D(D+1)$ parameters.
Thus, the dimension of the statistical manifold is $\frac{1}{2}D(D+3)$.

Calculating the information distance between $p(x|\mu,C)$ and $p(x|\mu
+d\mu,C+dC)$ is a matter of keeping track of all the indices involved.
Skipping all details, the result is%
\begin{equation}
d\ell^{2}=g_{ij}d\mu^{i}d\mu^{j}+g_{k}^{ij}dC_{ij}d\mu^{k}+g^{ij\,kl}%
dC_{ij}dC_{kl}~,
\end{equation}
where
\begin{equation}
g_{ij}=C_{ij}\,,\quad g_{k}^{ij}=0\,,\quad\text{and}\quad g^{ij\,kl}=\frac
{1}{4}(C^{ik}C^{jl}+C^{il}C^{jk})~,
\end{equation}
where $C^{ik}$ is the correlation matrix, that is, $C^{ik}C_{kj}=\delta
_{j}^{i}$. Therefore,
\begin{equation}
d\ell^{2}=C_{ij}dx^{i}dx^{j}+\frac{1}{2}C^{ik}C^{jl}dC_{ij}dC_{kl}~.
\label{metric Gaussian}%
\end{equation}

To conclude we consider a couple of special cases. For Gaussians that differ
only in their means the information distance between $p(x|\mu,C)$ and
$p(x|\mu+d\mu,C)$ is obtained setting $dC_{ij}=0$, that is,
\begin{equation}
d\ell^{2}=C_{ij}dx^{i}dx^{j}~,
\end{equation}
which is an instance of eq.(\ref{can3}). Finally, for spherically symmetric Gaussians,%

\begin{equation}
p(x|\mu,\sigma)=\frac{1}{(2\pi\sigma^{2})^{D/2}}\exp\left[  -\frac{1}%
{2\sigma^{2}}\delta_{ij}(x^{i}-\mu^{i})(x^{j}-\mu^{j})\right]  ~.
\end{equation}
The covariance matrix and its inverse are both diagonal and proportional to
the unit matrix,
\begin{equation}
C_{ij}=\frac{1}{\sigma^{2}}\delta_{ij}~,\quad C^{ij}=\sigma^{2}\delta
^{ij}\,,\quad\text{and}\quad c=\sigma^{-2D}~.
\end{equation}
Substituting
\begin{equation}
dC_{ij}=d\frac{1}{\sigma^{2}}\delta_{ij}=-\frac{2\delta_{ij}}{\sigma^{3}%
}d\sigma
\end{equation}
into eq.(\ref{metric Gaussian}), the induced information metric is
\begin{equation}
d\ell^{2}=\frac{1}{\sigma^{2}}\delta_{ij}d\mu^{i}d\mu^{j}+\frac{1}{2}%
\sigma^{4}\delta^{ik}\delta^{jl}\frac{2\delta_{ij}}{\sigma^{3}}d\sigma
\frac{2\delta_{kl}}{\sigma^{3}}d\sigma
\end{equation}
which, using
\begin{equation}
\delta^{ik}\delta^{jl}\delta_{ij}\delta_{kl}=\delta_{j}^{k}\delta_{k}%
^{j}=\delta_{k}^{k}=D~,
\end{equation}
simplifies to
\begin{equation}
d\ell^{2}=\frac{\delta_{ij}}{\sigma^{2}}d\mu^{i}d\mu^{j}+\frac{2D}{\sigma^{2}%
}(d\sigma)^{2}~.
\end{equation}

\section{Conclusion}

With the definition of the information metric we have only scratched the
surface. Not only can we introduce lengths and volumes but we can make use of
all sorts of other geometrical concepts such geodesics, normal projections,
notions of parallel transport, covariant derivatives, connections, and
curvature. The power of the methods of information geometry is demonstrated by
the vast number of applications. For a very incomplete point of entry to the
enormous literature in mathematical statistics see \cite{Amari Nagaoka
2000}\cite{Efron 1975}\cite{Rodriguez 1991}\cite{Kass Vos 1997}; in model
selection \cite{Myung et al 2000}\cite{Rodriguez 2006}; in thermodynamics
\cite{Ruppeiner 1995}; and for the extension to a quantum information geometry
see \cite{Balian et al 1986}\cite{Streater 1996}.

The ultimate range of these methods remains to be explored. In this tutorial
we have argued that information geometry is a natural and inevitable tool for
reasoning with incomplete information. One may perhaps conjecture that to the
extent that science consists of reasoning with incomplete information, then we
should expect to find probability, and entropy, and also geometry in all
aspects of science. Indeed, I would even venture to predict that once we
understand better the physics of space and time we will find that even that
old and familiar first geometry --- Euclid's geometry for physical space ---
will turn out to be a manifestation of information geometry. But that is work
for the future.


\begin{thebibliography}{99}                                                                                               %


\bibitem {Einstein 1949}A. Einstein, p. 67 in \textquotedblleft\emph{Albert
Einstein: Philosopher-Scientist}\textquotedblright, ed. by P. A. Schilpp (Open
Court 1969).

\bibitem {Cencov 1981}N. N. \v{C}encov: \emph{Statistical Decision Rules and
Optimal Inference}, Transl. Math. Monographs, vol. 53, Am. Math. Soc.
(Providence, 1981).

\bibitem {Amari 1985}S. Amari, \emph{Differential-Geometrical Methods in
Statistics} (Springer-Verlag, 1985).

\bibitem {Amari Nagaoka 2000}S. Amari and H. Nagaoka, \emph{Methods of
Information Geometry} (Am. Math. Soc./Oxford U. Press, 2000).

\bibitem {Caticha 2012}A. Caticha, \emph{Entropic Inference and the
Foundations of Physics} (USP Press, S\~{a}o Paulo, Brazil 2012); online at http://www.albany.edu/physics/ACaticha-EIFP-book.pdf.

\bibitem {Wootters 1981}W. K. Wootters, \textquotedblleft Statistical distance
and Hilbert space\textquotedblright, Phys. Rev. \textbf{D}, 357 (1981).

\bibitem {Balasubramanian 1997}V. Balasubramanian, \textquotedblleft
Statistical inference, Occam's razor, and statistical mechanics on the space
of probability distributions\textquotedblright, Neural Computation \textbf{9},
349 (1997).

\bibitem {Rao 1945}C. R. Rao, \textquotedblleft Information and the accuracy
attainable in the estimation of statistical parameters\textquotedblright,
Bull. Calcutta Math. Soc. \textbf{37}, 81 (1945).

\bibitem {Atkinson Mitchell 1981}C. Atkinson and A. F. S. Mitchell,
\textquotedblleft Rao's distance measure\textquotedblright, Sankhy\={a}
\textbf{43}A, 345 (1981).

\bibitem {Fisher 1925}R. A. Fisher, \textquotedblleft Theory of statistical
estimation\textquotedblright, Proc. Cambridge Philos. Soc. \textbf{122}, 700 (1925).

\bibitem {Rodriguez 1989}C. C. Rodr\'{\i}guez, \textquotedblleft The metrics
generated by the Kullback number\textquotedblright,\ \emph{Maximum Entropy and
Bayesian Methods}, J. Skilling (ed.) (Kluwer, Dordrecht 1989).

\bibitem {Campbell 1986}L. L. Campbell, \textquotedblleft An extended
\v{C}encov characterization of the information metric\textquotedblright%
,\ Proc. Am. Math. Soc. \textbf{98}, 135 (1986).

\bibitem {Efron 1975}B. Efron, Ann. Stat. \textbf{3}, 1189 (1975).

\bibitem {Rodriguez 1991}C. C. Rodr\'{\i}guez, \textquotedblleft Entropic
priors\textquotedblright,\ \emph{Maximum Entropy and Bayesian Methods}, edited
by W. T. Grandy Jr. and L. H. Schick (Kluwer, Dordrecht 1991).

\bibitem {Kass Vos 1997}R. A. Kass and P. W. Vos, \emph{Geometric Foundations
of Asymptotic Inference} (Wiley, 1997).

\bibitem {Myung et al 2000}J. Myung, V. Balasubramanian, and M.A. Pitt, Proc.
Nat. Acad. Sci. \textbf{97}, 11170 (2000).

\bibitem {Rodriguez 2006}C. C. Rodr\'{\i}guez, \textquotedblleft The ABC of
model selection: AIC, BIC and the new CIC\textquotedblright,\ \emph{Bayesian
Inference and Maximum Entropy Methods in Science and Engineering}, ed. by K.
Knuth \emph{et al.}, AIP Conf. Proc. Vol. \textbf{803}, 80 (2006)\ (omega.albany.edu:8008/CIC/me05.pdf).

\bibitem {Ruppeiner 1995}G. Ruppeiner, Rev. Mod. Phys. \textbf{67}, 605 (1995).

\bibitem {Balian et al 1986}R. Balian, Y. Alhassid and H. Reinhardt, Phys
Rep., \textbf{131}, 2 (1986).

\bibitem {Streater 1996}R. F. Streater, Rep. Math. Phys., \textbf{38}, 419-436 (1996).
\end{thebibliography}
\end{document}